\documentclass[aps,prl,print,groupedaddress,twocolumn,amsmath,amssymb,superscriptaddress]{revtex4-1}
\usepackage{graphicx}
\usepackage{dcolumn}
\usepackage{bm}
\usepackage{subfigure}
\usepackage{color}
\usepackage{txfonts}
\usepackage{natbib}

\begin{document}

\title{Persistent Josephson tunneling between Bi$_2$Sr$_2$CaCu$_2$O$_{8+x}$ flakes twisted by 45$^\circ$ across the superconducting dome}

\author{Yuying Zhu}\thanks{These authors contributed equally to this work.}
\affiliation{Beijing Academy of Quantum Information Sciences, Beijing 100193, China.}
\affiliation{Hefei National laboratory, Hefei 230088, China}
\author{Heng Wang}\thanks{These authors contributed equally to this work.}\affiliation{State Key Laboratory of Low Dimensional Quantum Physics and Department of Physics, Tsinghua University, Beijing, 100084, China.}

\author{Zechao Wang}\thanks{These authors contributed equally to this work.}
\affiliation{National Center for Electron Microscopy in Beijing, School of Materials Science and Engineering, Key Laboratory of Advanced Materials (MOE), The State Key Laboratory of New Ceramics and Fine Processing, Tsinghua University, Beijing 100084, China}
\affiliation{Ji Hua Laboratory, Foshan, Guangdong 528200, China}
\author{Shuxu Hu}\affiliation{State Key Laboratory of Low Dimensional Quantum Physics and Department of Physics, Tsinghua University, Beijing, 100084, China.}
\author{Genda Gu}\affiliation{Condensed Matter Physics and Materials Science Department, Brookhaven National Laboratory, Upton, New York 11973, USA}
\author{Jing Zhu}
\email{jzhu@mail.tsinghua.edu.cn}
\affiliation{National Center for Electron Microscopy in Beijing, School of Materials Science and Engineering, Key Laboratory of Advanced Materials (MOE), The State Key Laboratory of New Ceramics and Fine Processing, Tsinghua University, Beijing 100084, China}
\affiliation{Ji Hua Laboratory, Foshan, Guangdong 528200, China}
\author{Ding Zhang}
\email{dingzhang@mail.tsinghua.edu.cn}\affiliation{Beijing Academy of Quantum Information Sciences, Beijing 100193, China.}
\affiliation{State Key Laboratory of Low Dimensional Quantum Physics and Department of Physics, Tsinghua University, Beijing, 100084, China.}
\affiliation{RIKEN Center for Emergent Matter Science (CEMS), Wako, Saitama 351-0198, Japan}

\author{Qi-Kun Xue}
\email{qkxue@mail.tsinghua.edu.cn}\affiliation{Beijing Academy of Quantum Information Sciences, Beijing 100193, China.}
\affiliation{State Key Laboratory of Low Dimensional Quantum Physics and Department of Physics, Tsinghua University, Beijing, 100084, China.}
\affiliation{Southern University of Science and Technology, Shenzhen 518055, China.}
\date{\today}
\begin{abstract}
   There is a heated debate on the Josephson effect in twisted Bi$_2$Sr$_2$CaCu$_2$O$_{8+x}$ flakes. Recent experimental results suggest the presence of either anomalously isotropic pairing or exotic $d$+i$d$-wave pairing, in addition to the commonly believed $d$-wave one. Here, we address this controversy by fabricating ultraclean junctions with uncompromised crystalline quality and stoichiometry at the junction interfaces. In the optimally doped regime, we obtain prominent Josephson coupling (2-4 mV) in multiple junctions with the twist angle of 45$^\circ$, in sharp contrast to a recent report that shows two orders of magnitude suppression around 45$^\circ$ from the value at 0$^\circ$. We further extend this study to the previously unexplored overdoped regime and observe pronounced Josephson tunneling at 45$^\circ$ together with Josephson diode effect up to 50~K. Our work helps establish the persistent presence of an isotropic pairing component across the entire superconducting phase diagram.
\end{abstract}

\maketitle
The pairing symmetry of cuprate superconductors is of vital importance for constructing a microscopic theory of high temperature superconductivity~ \cite{Keimer2015,Tsuei2000,Zhong2016,Zhu_2021,Fan2021}. Among the proposals for verifying the $d$-wave pairing, an approach is to measure the phase-sensitive Josephson coupling between two twisted cuprates along the $c$-axis [Fig.~1(a)]~\cite{Klemm2005,Yokoyama2007,Arnold2000,Bille2001,Klemm2003}. Based on symmetry arguments, the Josephson tunneling between two $d$-wave superconductors vanishes when the twist angle along $c$-axis is 45$^\circ$, while a strongest tunneling is expected when the twist angle is 0$^\circ$. In the case of $s$-wave superconductors, the Josephson current persists to flow with a constant value, independent of the twist angle. Apart from the two scenarios, it was recently proposed~\cite{Can2021,Tummuru2022,Zhao2021} that co-tunneling of Cooper pairs may occur between two cuprate monolayers at the twist angle around 45$^\circ$, giving rise to an emergent $d$+i$d$ or $d$+i$s$ wave pairing. Whether this exotic scenario, which promises topological superconductivity, is applicable to a strongly correlated system remains to be verified~\cite{Song2022,Lu2022}.

The intriguing proposals above call for careful experimental tests, which can be carried out by using Bi$_2$Sr$_2$CaCu$_2$O$_{8+x}$(BSCCO) crystals. BSCCO naturally consists of a series of intrinsic Josephson junctions (IJJ) along the $c$-axis~\cite{Kleiner1992,Kleiner1994} and it can be mechanically cleaved~\cite{Liao2018,Yu2019} into two parts and re-assembled after twisting one of them with a predefined angle. This re-assembling does not impose extra strain or induce changes in stoichiometry, allowing for the realization of atomically flat interface as the tunnel barrier. By contrast, the in-plane Josephson junctions realized by film growth suffer from large structural distortion, tunneling plane misalignment and chemical inhomogeneity at the grain boundary~\cite{Tsuei2000,Zhang1996,Jin2002}. Historically, the $c$-axis twisted BSCCO junctions were realized experimentally by using bulk bicrystals~\cite{Li1999} and whiskers~\cite{Takano2002,Latyshev2004}. These experiments favored isotropic pairing instead of the $d$-wave pairing~\cite{Klemm2005,Arnold2000,Bille2001,Klemm2003}. However, it remains unclear if these macroscopic junctions maintained the atomically sharp and uniform interface with uncompromised crystalline quality~\cite{Tsuei2000}. There also exist technical issues such as overheating in bulk samples ~\cite{Li1999} and participation of multiple intrinsic junctions~\cite{Takano2002,Latyshev2004}. Recently, micrometer-sized BSCCO junctions were realized by the van~der~Waals (vdW) stacking technique. The atomically flat interface was revealed to extend over the complete junction area~\cite{Zhu_2021}.

\begin{figure}
\includegraphics[width=86mm]{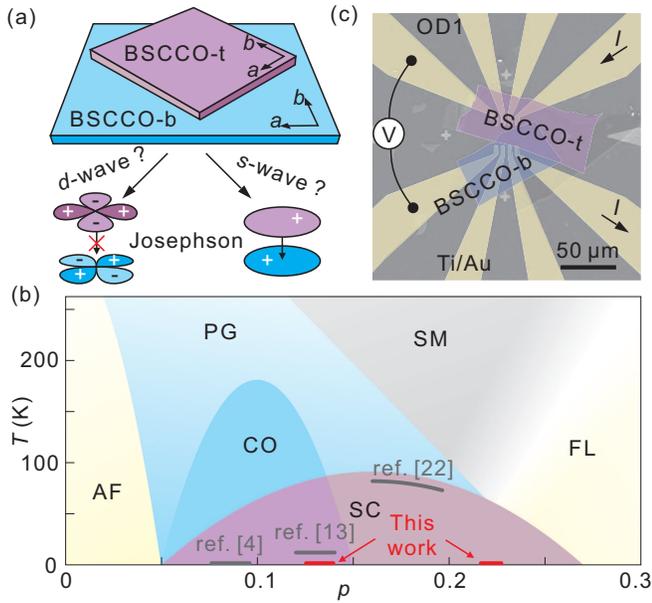}
\caption{\label{fig1}
  (a)~Schematic illustration of the twisted BSCCO junction and the theoretical expectation of the Josephson tunneling based on different pairing symmetries. Here the top BSCCO (BSCCO-t) is rotated against the bottom BSCCO (BSCCO-b) by 45$^\circ$. (b)~Typical phase diagram of high temperature cuprate superconductors including antiferromagnetism (AF), pseudo-gap (PG), charge ordering (CO), strange metal phase (SM), Fermi liquid (FL) and superconductivity (SC). Solid gray curves indicate the doping and temperature range of previous studies~\cite{Zhu_2021,Zhao2021,Li1999}.  (c)~False-colored SEM image of a twisted junction (sample OD1).
  }
\end{figure}

However, there is still a lack of consensus on the angular dependence of the Josephson coupling strength, defined as the product the Josephson critical current and the normal state resistance--$I_c R_n$. Twisted junctions in the underdoped (UD) regime ($p<0.1$) showed large $I_c R_n$ at 45$^\circ$~\cite{Zhu_2021}, indicative of isotropic pairing. Yet another experiment~\cite{Zhao2021} with samples in the nearly optimally doped (OP) regime reported suppression of $I_c R_n$ by two orders of magnitude as the twist angle varied from 0$^\circ$ to 45$^\circ$. The discrepancy requests further clarification. One remaining technical issue is that the interfaces often exhibit reduced signal in the atomically resolved image and expanded interlayer spacing~\cite{Lee_2021,Zhu_2021,Zhao2021}. Moreover, the tunneling experiments so far focused on the doping regime from UD to OP [Fig.~1(b)]. The phase diagram in this regime is complicated by the charge ordering, pseudogap and strange metal phases. By contrast, the phase diagram in the OD regime is simpler and the corresponding cuprates exhibit well-established Fermi surface~\cite{Vignolle2008,Yu2021,Zhao2022}. A recent theoretical study suggested a non-trivial change in the phase difference of the twisted junction as the doping level moves to the OD regime~\cite{Lu2022}. It is, therefore, necessary to extend the study to the OD regime for a comprehensive understanding of the tunneling phenomena.

In this Letter, we address the Josephson effect of twisted cuprate junctions in both OP and OD regimes [Fig.~1(b)]. We fabricate ultrathin twisted junctions of BSCCO flakes at 45$^\circ$ by an on-site cold stacking technique. By high resolution transmission electron microscopy (HR-TEM), we demonstrate that our junctions meet the demanding requirement: all the atoms at the interface possess the same signal intensity as those in the bulk, attesting to the uncompromised crystalline quality. Transport experiments on the same batch of junctions indicate: 1) preserved doping level at the interface as that of the bulk crystal; 2) strong Josephson tunneling with $I_c R_n$ as large as 4~mV for OP junctions and 2~mV for OD ones at the twist angle close to 45$^\circ$; 3) conventional temperature dependence of $I_c$, in disagreement with that predicted for $d$+i$d$-wave pairing. We further unveil the asymmetric tunneling observed in some junctions, showing Josephson diode effect up to about 50 K. Finally, we compare $I_c R_n$ at 45$^\circ$ and 0$^\circ$ in both OP and OD regimes and observe only weak angular dependence, indicating a prominent isotropic pairing component over the whole phase diagram.

\begin{figure}
\includegraphics[width=86mm]{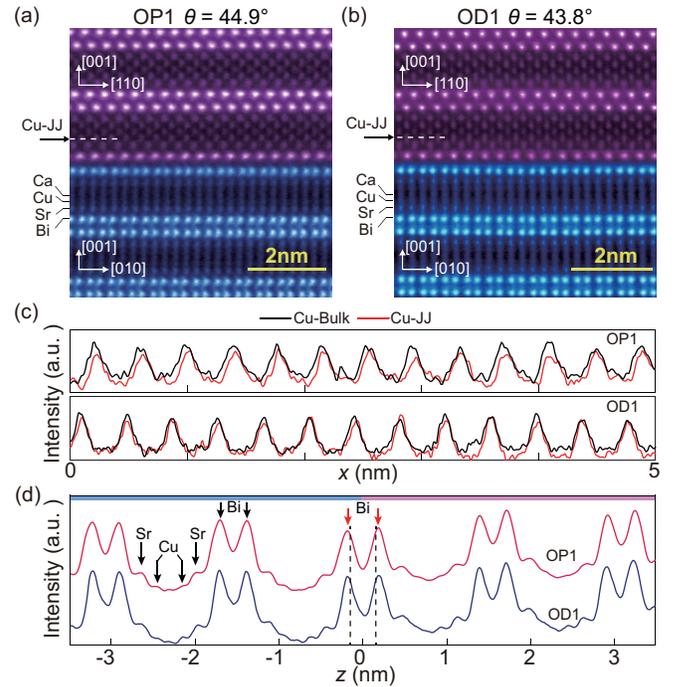}
\caption{\label{fig2}
  (a)(b)~Cross-sectional high-angle annular dark field scanning transmission electron microscopy (HAADF-STEM) images of four twisted BSCCO junctions. (c)~Intensity profile of the CuO$_2$ plane close to the interface [dashed lines in (a)(b)] and that in the bulk. (d)~Normalized integrated intensity profiles of the TEM images along the $c$-axis. Red arrows mark the peaks from BiO layers at the interface. Here, $z$ = 0 marks the twist boundary. Curves are horizontally offset for clarity.
  }
\end{figure}

\begin{figure}
\includegraphics[width=86mm]{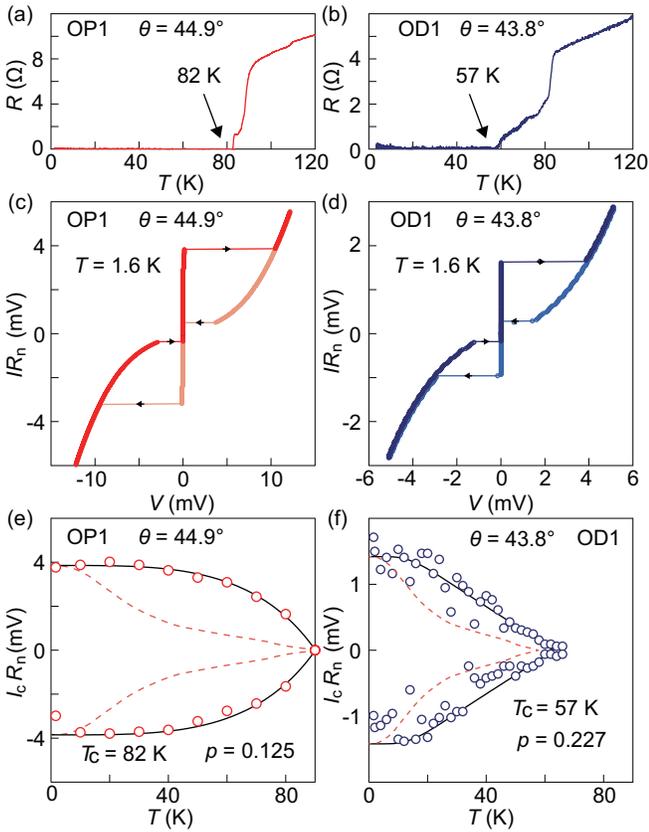}
\caption{\label{fig3}
  (a)(b)~Temperature-dependent resistance across the junctions of OP1 and OD1. The arrows mark $T_c$. (c)(d)~Normalized tunneling characteristics at $T$ = 1.6~K. Darker (lighter) color reflects data points taken in the positive (negative) sweeping direction, as indicated by the arrows. (e)(f)~Temperature dependence of $I_c R_n$. Solid curves are theoretical fits by using a modified Ambergaokar-Baratoff formula~\cite{Zhu_2021}. Dashed curves are the theoretical prediction based on $d$+i$d$ pairing~\cite{Tummuru2022}.
  }
\end{figure}

\begin{figure}
\includegraphics[width=86mm]{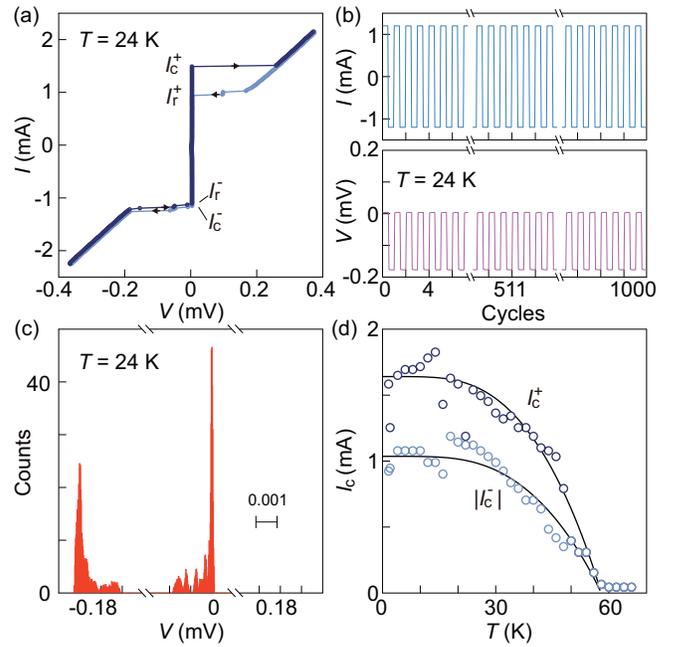}
\caption{\label{fig4}
  (a)~Current-voltage characteristics of OD-BSCCO junction with a nominal twist angle of 45$^\circ$ at $T$ = 1.6~K and 24~K. Dark and light colors indicate data obtained in two opposite sweeping directions, as indicated by the arrows. $I_c^\pm$ and $I_r^\pm$ represent the critical current and re-trapping current in the positive/negative sweeping directions. The data are obtained at nominally zero magnetic field. (b) Junction voltage (bottom) under the square wave of excitation current ($\pm$1.2~mA, 0.03~Hz) (top) at $T$ = 24~K. (c)~Histograms of the measured junction voltage data in (b). (d)~Temperature dependence of the critical Josephson currents in two directions. Solid curves are guide to the eye.
  }
\end{figure}

\begin{figure}
\includegraphics[width=86mm]{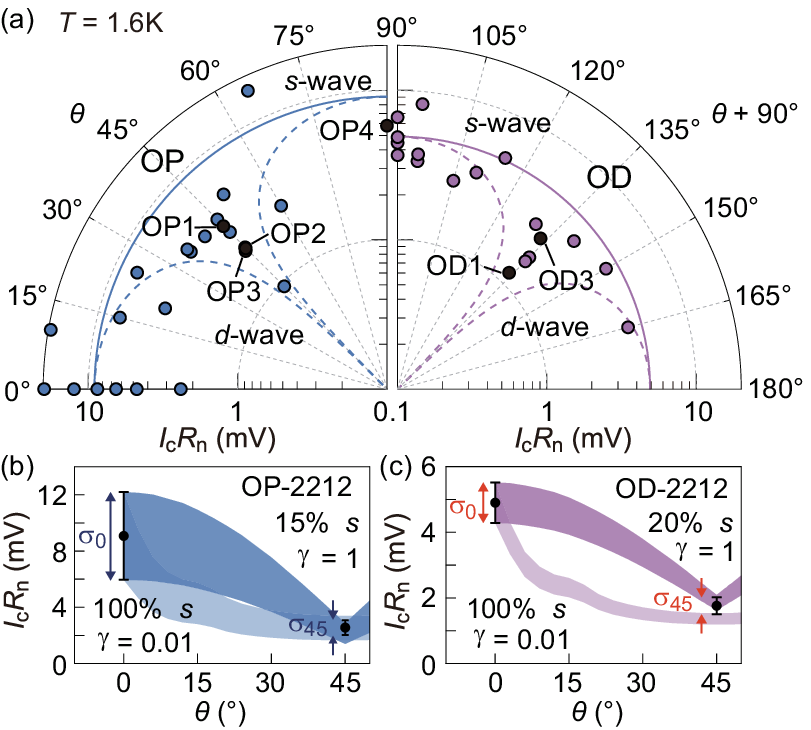}
\caption{\label{fig5}
  (a)~Fan-chart diagrams of $I_c R_n$ as a function of twist angles. Except for the filled symbols with angles determined by TEM, we use nominal twist angles. Upper-left/upper-right quadrant includes data points from junctions in the OP/OD doping regime. $T_c$ for OD is in the range of 57 to 67~K.(b)(c)~Comparison between the averaged $I_c R_n$ at 0$^\circ$ and that at 45$^\circ$ for OP and OD. Circles represent the averaged values and the error bars are the corresponding sampling standard deviations. The shaded bands with darker colors represent the theoretically expected behavior when the order parameter is a mixture of $s$-wave and $d$-wave. The percentage of $s$-wave component is indicated in the panels. The tunneling is considered to be incoherent as represented by factor $\gamma$. The upper and lower bounds of the band take into account the data scattering $\sigma_0$ at 0$^\circ$. The shaded bands with lighter colors are theoretically calculated angular dependence for the tunneling between two pure $s$-wave superconductors with enhanced coherence in tunneling ($\gamma$ = 0.01). Details of the calculation is given in the supplementaxry note IV and in ref.~\cite{Zhu_2021}.
  }
\end{figure}

We fabricate the Josephson junctions [an example is given in Fig. 1(c)] out of high quality single crystals of BSCCO~\cite{Gu1993} with the cold van der Waals stacking technique [Supplementary note I and Fig. S1]. The high crystalline quality of the twisted junctions are demonstrated by TEM. Figure 2(a) shows the representative images of two samples in the OP and OD regimes (OP1 and OD1). The samples have a nominal twist angle of 45$^\circ$ such that the cross-sectional images reveal the top and bottom BSCCO structure from different crystalline orientations. We determine the exact twist angles by using the Kikuchi patterns~\cite{Zhu_2021} [values are indicated in the respective TEM image]. In particular, the deviation from perfect 45$^\circ$ for OP1 and OP2 [see Fig. S2] is as small as 0.05-0.1$^\circ$. We denote the twist angles of them as 44.9$^\circ$ (45.1$^\circ$ is equivalent). As shown in Fig. 2(a), the bottom sections of BSCCO in the images show bunching of Bi atoms (brightest dots) in OP1/OD1. The artificially twisted interface is immediately discerned because only one of the double BiO layers there hosts bunching.

Images of equivalently superb quality are obtained from multiple samples and in horizontally displaced regions around the junction area (further examples in Figs.~S2,S3). From the TEM images, we observe that the atoms of Bi, Sr, Cu and Ca all show comparable intensities at the interface and in the bulk, demonstrating the uncompromised crystalline quality. For the superconducting layer, we take a horizontal line cut along the CuO$_2$ plane of the top BSCCO [indicated by arrows in Fig.~2(a)(b)] next to the twist boundary. We compare this line profile with that obtained from the CuO$_2$ plane far away from the interface, as shown in Fig.~2(c). Each peak represents one row of Cu atoms. Clearly, the two line profiles are closely matched. Fig.~2(d) shows the averaged intensity profile along the $c$-axis. It demonstrates that the double BiO planes at the twist boundary (indicated by red arrows) exhibit intensity peaks comparable to those from the bulk (indicated by black arrows). In general, the artificial interface is atomically flat without any reconstruction or wrinkles~\cite{Zhu_2021}. In addition, we point out that the thickness of the artificial junctions is slightly larger than that of IJJ. The separation between the vertical dashed lines in Fig.~2(d) represent the distance between nearest neighbored BiO layers in the bulk. We observe that the actual peaks from the BiO plane at the twist boundary (red arrows) situate away from the dashed lines. Quantitatively, the distance between the two BiO planes on average is larger than the bulk value by 12$\%$ (Fig.~S3). Apart from this slightly thicker tunnel barrier, the twisted junction has a smaller tunneling area$\--$the overlapping region between the top and bottom flakes$\--$than the IJJ in either the top or bottom BSCCO [Fig.~1(d)]. These features help distinguish the transport at the twisted interface from that of IJJ, as discussed below.

Figure~3 shows the transport results from OP1 and OD1. Similar results from other samples are given in Figs. S2, S4 and S5. For the OP samples, temperature dependent resistance measurements indicate a narrow superconducting transition with $T_c$ close to that of the bulk [Fig.~3(a), Fig.~S2(c) and Fig. S4], confirming the preserved doping. For the OD samples [Fig.~3(b), Fig.~S2(d) and Fig. S5], however, the junction resistance reaches zero at around 50~K but the onset for the superconducting transition starts at a higher temperature of about 80~K. We emphasize that the OD doping level is preserved at the twisted interface, because the junction is formed at a cryogenic temperature (-50~$^\circ$C) that suppresses oxygen out-diffusion and is further buried inside the relatively thick stack of BSCCO. We attribute the higher onset temperature to the top surface of OD-BSCCO flakes, which suffers from oxygen loss in the final fabrication step at a relatively alleviated temperature (-30~$^\circ$C). Further support to this argument is given in the supplementary information (Supplementary note III and Fig. S6). The OP samples do not suffer from this problem presumably because the out-diffusion is only prominent in the superoxygenated state.

Figure~3(c)(d) show the Josephson tunneling characteristics. We normalize the current by the normal state resistance ($R_n$) for a better comparison among samples. In each figure, the single vertical branch at zero bias reflects the Josephson effect between the two twisted cuprate layers. The IJJ do not contribute here because their critical currents are much higher$\--$due to the larger tunneling area and relatively thinner barrier$\--$and are not reached in our measurement. The vertical bar essentially represents the Josephson coupling strength--$I_c R_n$. Notably, we obtain $I_c R_n$ (44.9$^\circ$) = 4~mV at OP. This is in sharp contrast to the previous report in the same doping regime, which shows that $I_c R_n$ (44.9$^\circ$) was as small as 0.19~mV. This difference may indicate that there exists statistical fluctuation among samples because so far only a few samples are reported to be within 45.0$\pm$0.1$^\circ$ in OP (OP1 and OP2 in our case and one in ref.~\cite{Zhao2021}). Differences in the fabrication steps are given in the supplementary material.

Next, we show that our results are inconsistent with the $d$+i$d$ or $d$+i$s$-wave pairing scenario~\cite{Tummuru2022}. Such an emergent pairing can give rise to Josephson tunneling of paired Cooper pairs at the twist angles where pure $d$-wave pairing demands strong suppression at 45$^\circ$. The dashed curves in Figure~3(e)(f) indicate the predicted temperature dependence for this co-tunneling process~\cite{Tummuru2022}. In sharp contrast to this behavior, our experimental results show quite standard behaviors similar to that prescribed by the Ambegaokar-Baratoff (A-B) formula~\cite{Ambegaokar1963}. We also comment on the non-monotonic temperature dependence at twist angles away from $45^\circ$, which was also predicted to be a manifestation of exotic pairing~\cite{Tummuru2022}. In experiment, we indeed observe non-monotonic or non-standard temperature dependence (Fig.~S4 and S5) but the non-monotonic behavior even appears at $0^\circ$ [Fig.~S5(a)]. It indicates that the temperature dependence may be influenced by other extrinsic factors. In particular, we speculate that unintentional flux trapping may be the driving mechanism, since suppression of $I_c R_n$ at low temperatures was observed in the intrinsic Josephson junctions under a small magnetic field~\cite{Yurgens_1999}.

Interestingly, the trapped fluxes are found to give rise to an asymmetric tunneling in some devices. In Fig.~4, we provide a typical example of such a Josephson diode effect~\cite{Wu_2022}. Figures~4(a) show the $I$-$V$ characteristics of an OD junction with $\theta_{\rm{nominal}}=45^\circ$. The Josephson critical current in the positive axis is obviously larger than the absolute value in the negative axis, i.e. $I_c^+>\left|I_c^-\right|$. The asymmetry factor $\eta=\frac{(I_c^+ - \left|I_c^-\right|)}{(I_c^+ + \left|I_c^-\right|)}$ is as high as 26\% at 1.6~K. This asymmetry is quite stable and the rectification effect over one thousand repetitions is achieved, as shown in Fig.~4(b). Figure~4(c) displays the histogram of the measured voltages in Fig.~4(b). With an input square wave of current jumping between 1.2~mA and -1.2~mA, the output voltage is either -0.18~mV or 0~mV. Figure~4(d) further demonstrates that the asymmetric effect persists over the entire superconducting regime up to 50~K. Although a 45$^\circ$-twist junction is used here, similar behaviors can be found at $\theta=0^\circ$ as well (Fig.~S7). We note that the results in Fig.~4 are obtained without applying the magnetic field. However, we obtain the peak-dip structure of $\eta$ as a function of $B$ with its center away from $B=0$~T (Fig.~S7). This observation clearly indicates that there exists a remanent field, which is approximately 1~mT. The driving mechanism for the Josephson diode effect remains unknown at this stage~\cite{Ghosh_2022}. Nevertheless, our experiments suggest that a carefully designed magnetic field shield must be applied in order to address the non-standard temperature dependence of $I_c$ due to intrinsic properties.

In Fig.~5(a), we provide an overview of our data points as a function of twist angles at both OD and OP levels. In each doping regime, we observe appreciable $I_c R_n$ in multiple samples around 45$^\circ$. These data points apparently fall outside the expected behavior of a pure $d$-wave pairing symmetry. However, $I_c R_n$ in OP and OD indeed shows a drop as $\theta$ increases from 0$^\circ$ to 45$^\circ$, if comparing the maximal values obtained at different angles. This decrease can be further appreciated in Fig.~5(b)(c), where the averaged values at 45$\pm$1$^\circ$ for OP or 45$\pm$2$^\circ$ for OD with those at 0$^\circ$ are compared. On the one hand, this angular dependence can be explained by a mixture of isotropic $s$-wave and anisotropic $d$-wave components~\cite{Zhu_2021}. We show the expected angular behavior for a mixture of 15$\%$/20$\%$ $s$-wave and 85$\%$/80$\%$ $d$-wave as shaded stripes (darker colors), of which the upper and lower bounds are set by the sampling standard deviation at 0$^\circ$. Such a mixture can yield the expected value at 45$^\circ$ that is in agreement with our data. It suggests the existence of a substantial portion of isotropic pairing (15-20$\%$), comparable to that in YBCO (15$\%$)~\cite{Smilde2005}. On the other hand, the decreasing trend can be solely attributed to the orbital effect. To further illustrate this point, we consider tunneling between two 100$\%$ $s$-wave superconductors with tunneling coherence that is 100 times higher than that in the $s$/$d$ mixed situation~\cite{Klemm2005}. The shaded bands with lighter colors in Fig.~5(b)(c) show the calculated behaviors, which also reproduces the suppressed $I_c R_n$ at 45$^\circ$ observed in experiment. In reality, the two scenarios discussed above$\--$$s$-wave/$d$-wave mixing and orbital effect$\--$may both participate in giving rise to the angular dependence seen in experiment. While we cannot rule out the $d$-wave pairing scenario for the time being, the estimated 15-20$\%$ of $s$-wave represents at least the lower bound.

In summary, we fabricate twisted Josephson junctions at OD and OP doping levels, and demonstrate that they are of unprecedentedly high crystalline quality at the interfaces. Low-temperature transport reveals strong Josephson tunneling at the twist angle of ~45$^\circ$. The conventional temperature dependence of these twisted junctions speaks against the presence of $d$+i$d$ or $d$+i$s$-wave pairing. Intriguingly, we observe Josephson diode effect in the junctions, setting the temperature for observing such an effect to a record high value of tens of kelvin. From the angular dependence of the Josephson coupling, we conclude that there exists an indispensable isotropic pairing component in the twisted Josephson junctions.

\begin{acknowledgments}
This work is financially supported by the National Natural Science Foundation of
China (grants No. 51788104, No. 12141402, No. 12004041, No. 11922409, No. 11790311,
5); Ministry of Science and
Technology of China (2017YFA0302902, 2017YFA0304600); Innovation Program for Quantum Science and Technology (Grant No. 2021ZD0302600).
\end{acknowledgments}

%


\end{document}